\documentclass[reprint,superscriptaddress,nofootinbib,amsmath,amssymb,aps,prl,floatfix,]{revtex4-2}

\usepackage{graphicx}
\usepackage{dcolumn}
\usepackage{bm}
\usepackage{xcolor}
\usepackage{soul}
\usepackage[
colorlinks=true,        
citecolor=blue,         
linkcolor=blue,         
urlcolor=blue           
]{hyperref}             

\usepackage{booktabs} 

\newcommand{\Eq}[1]{Eq.~\eqref{#1}}     
\newcommand{\Fig}[1]{Fig.~\ref{#1}}     
\newcommand{\Table}[1]{Table~\ref{#1}}  

\newcommand{\dif}{\mathrm{d}}

\begin{document}

\title{Search for stochastic gravitational-wave background from string cosmology with Advanced LIGO and Virgo's O1$\sim$O3 data}

\author{Yang Jiang}
\email{jiangyang@itp.ac.cn} 
\affiliation{CAS Key Laboratory of Theoretical Physics, 
		Institute of Theoretical Physics, Chinese Academy of Sciences,
		Beijing 100190, China}
\affiliation{School of Physical Sciences, 
		University of Chinese Academy of Sciences, 
		No. 19A Yuquan Road, Beijing 100049, China}

\author{Xi-Long Fan}	\email{Corresponding author:xilong.fan@whu.edu.cn}
\affiliation{School of Physics and Technology, Wuhan University, Wuhan, Hubei 430072, China}
		
\author{Qing-Guo Huang}
\email{Corresponding author: huangqg@itp.ac.cn}
\affiliation{CAS Key Laboratory of Theoretical Physics, 
		Institute of Theoretical Physics, Chinese Academy of Sciences,
		Beijing 100190, China}
\affiliation{School of Physical Sciences, 
		University of Chinese Academy of Sciences, 
		No. 19A Yuquan Road, Beijing 100049, China}
\affiliation{School of Fundamental Physics and Mathematical Sciences
		Hangzhou Institute for Advanced Study, UCAS, Hangzhou 310024, China}

\date{\today}

\begin{abstract}
String cosmology models predict a relic background of gravitational-wave (GW) radiation in the early universe. The GW energy spectrum of radiated power increases rapidly with the frequency, and therefore it becomes a potential and meaningful observation object for high-frequency GW detector. We focus on the stochastic background generated by superinflation in string theory and search for such signal in the observing data of Advanced LIGO and Virgo O1$\sim$O3 runs in a  Bayesian framework. We do not find the existence of the signal, and thus put constraints on the GW energy density. Our results indicate that at $f=100\,\text{Hz}$, the fractional energy density of GW background is less than $1.7\times10^{-8}$ and $2.1\times10^{-8}$ for dilaton-string and dilaton only cases respectively, and further rule out the parameter space restricted by the model itself due to the non-decreasing dilaton and stable cosmology background ($\beta$ bound).
\end{abstract}
\maketitle

\section{Introduction}
The standard cosmological model \cite{Coley:2019yov} has achieved great success in describing the behavior of our Universe. Combined with the inflationary period \cite{PhysRevD.23.347}, such a model provides a natural way to understand the puzzles like fine-tuning initial conditions and gives good agreements with the inhomogeneous structure we have observed. Nevertheless, this mechanism is still far from perfect. In most models of inflation based on a scalar field coupling to gravity minimally, it lasts so long that the physical fluctuation corresponding to present large-scale structure will shrink to the scale smaller than Planck length at the beginning of inflation. This is the so-called ``trans-Planckian" problem \cite{PhysRevD.63.123501}. Moreover, the spacetime curvature also increases when we retrospect in time and as a result,  we meet initial singularity \cite{PhysRevLett.72.3305,PhysRevLett.90.151301} from big bang inevitably. Likewise, now we know little about the physics essence of the inflation field due to its exotic characters.

Quantum effect of gravity is inevitable in the primordial Universe. So the proposals dealing with such problems may be found in string theory. The resulting string cosmology leads to a pre-big bang scenario \cite{Gasperini:1992em,Gasperini:2007vw} where extra dimensions are introduced and small characteristic size of string \cite{PhysRevLett.55.1036} avoids the initial singularity in general relativity. The Universe can start inflation with a large Hubble horizon, also solving the trans-Planckian problem. String theory predicts the existence of a scalar dilaton field coupling to gravity, and it evokes the inflation which is different from standard slow-roll inflation \cite{AndreiLinde_1991}. As an interesting consequence, pre-big bang inflation produces a primordial stochastic gravitational wave background (SGWB) which is blue titled \cite{PhysRevD.43.2566,Brustein:1995ah}. That means an power spectrum increasing with frequency. Therefore, ground based interferometers like  Advanced LIGO \cite{AdvancedLIGO2015} and Virgo \cite{VIRGO:2014yos} play an important role in verifing and constraining the parameters of the pre-big bang model. 

The spectral density of such SGWB was considered in \cite{PhysRevD.47.1519,Brustein:1995ah,Brustein:1994kn} and the detection prospect was discussed in \cite{Gasperini:2016gre}. The main conclusion states that although there have been constraints from cosmic mircowave background (CMB) observations, big-bang nucleosynthesis (BBN), millisecond pulsar timing \cite{Kaspi1994HighP}, etc,
there is still a wide allowed parameter space left to be detectable with the increase of the sensitivity of ground detectors. Besides, the detecting ability of  parameter space for string cosmology have been investigated  using simulated noise curves  \cite{1997PhRvD..55.3260A,2008PhLB..663...17F,2019ApJ...887...28L} with Neyman-
Pearson criterion.

In 2015, the successful detection of compact binary coalescence GW150914 \cite{LIGOScientific:2016emj} was quite inspiring and marked the beginning of gravitational wave (GW) astronomy. Besides, the LIGO/Virgo/KAGRA scientific collaboration (LVK) has been devoted to finding a general stochastic background. Until O3 observing run, there is no SGWB detected. Therefore, only upper limits on its energy density can be put \cite{KAGRA:2021kbb}. In this letter, for the first time,  we adopt the observing data from O1 to O3 runs \cite{LIGOScientific:2016jlg,LIGOScientific:2019vic,KAGRA:2021kbb,LVK:IsoSGWBdata} and search for the GWs signal from string cosmology.

\section{SGWB from string cosmology}
The fractional energy density of SGWB is depicted as
\begin{equation}
    \Omega_\text{gw}(f)=\frac{8\pi G}{3c^2 H_0^2}\frac{\dif \rho_\text{gw}}{\dif \ln f},
\end{equation}
where $\rho_\text{gw}$ is the energy density of GWs between $f$ and $f+\dif f$. In the string cosmology scenario, there are many models in which the Universe undergone a dilaton-driven inflation followed by a string epoch. Both electromagnetic radiation and GWs are emitted during these periods, but GWs decoupled earlier and it can more truly reflect the Universe. Then after a possible short dilaton-relaxation era, the evolution came into standard radiation dominated cosmology. In this letter, we use a simple model to approximate the spectrum \cite{Brustein:1996mq}
\begin{equation}
    \Omega_\text{gw}(f) = \begin{cases}
    \Omega_\text{gw}^s (f/f_s)^3, & f<f_s \\
    \Omega_\text{gw}^s (f/f_s)^\beta, & f_s<f<f_1 \\
    0, & f_1<f
    \end{cases}.
    \label{eq:spectrum}
\end{equation}
$f_s$ and $\Omega_\text{gw}^s$ are frequency and fractional energy density produced at the end of the dilaton-driven inflation phase. $\beta$ represents the logarithmic slope of the spectrum produced during the string epoch and it equals to
\begin{equation}
    \beta = \frac{\log \left(\Omega_\text{gw}^\text{max}/\Omega_\text{gw}^s \right)}{\log \left( f_1/f_s\right)}.
\end{equation}
$f_1$ is the maximum frequency above which GW is not produced. Following \cite{PhysRevD.55.3260}, we set the cut-off frequency
\begin{equation}
    f_1 = 1.3\times10^{10} \left( \frac{H_r}{5\times10^{17}\,\text{GeV}}\right)^{1/2}\,\text{Hz}, \label{eq:f_1}
\end{equation}
and corresponding energy density occurs at $f_1$
\begin{equation}
    \Omega_\text{gw}^\text{max} = 1\times10^{-7} h_{100}^{-2}\left( \frac{H_r}{5\times10^{17}\,\text{GeV}}\right)^{2}. \label{eq:omega_max}
\end{equation}
$H_r$ is the Hubble parameter when the string epoch ends. We set the reduced Hubble constant $h_{100}=0.679$ in the analysis.

In fact, $\beta$ is related to the basic parameters of string cosmology models. To go beyond the unknown arguments, let $\Omega_\text{gw}$ equals to zero when $f>f_S$, which means that there is no GW emitted in string phase. This phenomenological model is so-called ``dilaton only" case and the spectrum becomes
\begin{equation}
    \Omega_\text{gw}(f) = \begin{cases}
    \Omega_\text{gw}^s (f/f_s)^3, & f<f_s \\
    0, & f_s<f
    \end{cases}.
\end{equation}

\section{Data analysis}
SGWB will cause a cross-correlation between a pair of detectors. Within the strain data $\tilde{s}_{I,J}$ from the detectors, we can construct an estimator
\begin{equation}
     \hat{C}_{IJ}(f)=\frac{2}{T}\frac{\text{Re}[\tilde{s}_I^*(f)\tilde{s}_J(f)]}{\gamma_{IJ}(f)S_0(f)}, \label{eq:correlation}
\end{equation}
where $\gamma_{IJ}(f)$ is the overlap reduction function \cite{PhysRevD.48.2389} of this baseline $IJ$. Its magnitude denotes the sensitivity of the detector pairs. $S_0(f)=(3H_0^2)/(10\pi^2f^3)$ is a normalized factor. $T$ is the observing time. In the absence of correlated noise, \Eq{eq:correlation} is normalized as $\langle \hat{C}_{IJ}(f) \rangle=\Omega_\text{gw}(f)$ and its variance is
\begin{equation}
    \sigma_{IJ}^2=\frac{1}{2T\Delta f}\frac{P_I(f)P_J(f)}{\gamma_{IJ}^2(f)S_0^2(f)},
\end{equation}
where $P_{I,J}$ is the power spectral density of the detectors. In order to estimate parameters $\bm{\theta}$ related to certain model $\mathcal{M}$, we adopt Bayesian framework \cite{PhysRevLett.109.171102} to calculate the posterior probability
\begin{equation}
    p(\bm{\theta},\mathcal{M}|\hat{C})\propto p(\hat{C}|\bm{\theta})p(\bm{\theta},\mathcal{M}).
\end{equation}
$\hat{C}$ is tested to be Gaussian-distributed \cite{KAGRA:2021kbb}. For more than one baseline, the likelihoods are combined to be
\begin{equation}
    p(\hat{C}|\bm{\theta})\propto \exp\left[-\frac12\sum_{IJ}\sum_{f}\frac{\left(\hat{C}_{IJ}(f)-\Omega_\text{gw}(f;\bm{\theta})\right)^2}{\sigma^2_{IJ}(f)}\right].
\end{equation}
The ratio of evidences, so-called Bayes factor, between two hypotheses
\begin{equation}
    \mathcal{B}_{12}=\frac{p(\hat{C}|\mathcal{M}_1)}{p(\hat{C}|\mathcal{M}_2)}=\frac{\int p(\hat{C}|\bm{\theta}_1)p(\bm{\theta}_1,\mathcal{M}_1)\, \dif\bm{\theta}_1}{\int p(\hat{C}|\bm{\theta}_2)p(\bm{\theta}_2,\mathcal{M}_2)\, \dif\bm{\theta}_2},
\end{equation}
can be used to tell which models fit the observing results better. Dynamic nested sampling package \texttt{dynesty} is adopted to carry out the algorithm described above \cite{speagle2020dynesty}.
\begin{table}[ht]
    \centering
    \begin{tabular}{p{0.3\columnwidth}<{\centering}p{0.6\columnwidth}<{\centering}}
        \toprule
        Parameters & Priors \\
        \midrule
        $\Omega_\text{gw}^s$ & $\text{LogUniform}[10^{-13},\; 5\times10^{-5}]$ \\
        $f_s\,(\text{Hz})$ & $\text{LogUniform}[50,\; 300]$ \\
        $H_r\,(\text{GeV})$ & $\text{LogUniform}[10^{12},\; 10^{19}]$ \\
        \bottomrule
    \end{tabular}
    \caption{Prior distributions $p(\bm{\theta},\mathcal{M})$ for the parameters in the analysis.}
    \label{tab:priors}
\end{table}

$\bm{\theta}=(\Omega_\text{gw}^s,\, f_s,\, H_r)$ are free parameters to be determined. The priors we adopted are summarized in \Table{tab:priors}. We set a log uniform prior between $10^{-13}$ and $5\times10^{-5}$ to $\Omega_\text{gw}^s$. The selection of this lower bound is inherited from \cite{KAGRA:2021kbb} and a quite large upper bound is set to avoid the leakage of posterior. We place $f_s$ in the sensitive band of the detectors. And the prior of $H_r$ is choosed based on the cut-off frequency $f_1$ varying in $10^7\sim10^{11}$ Hz \cite{Galluccio:1996xa}.

\begin{figure}[ht]
    \centering
    \includegraphics[width=\columnwidth]{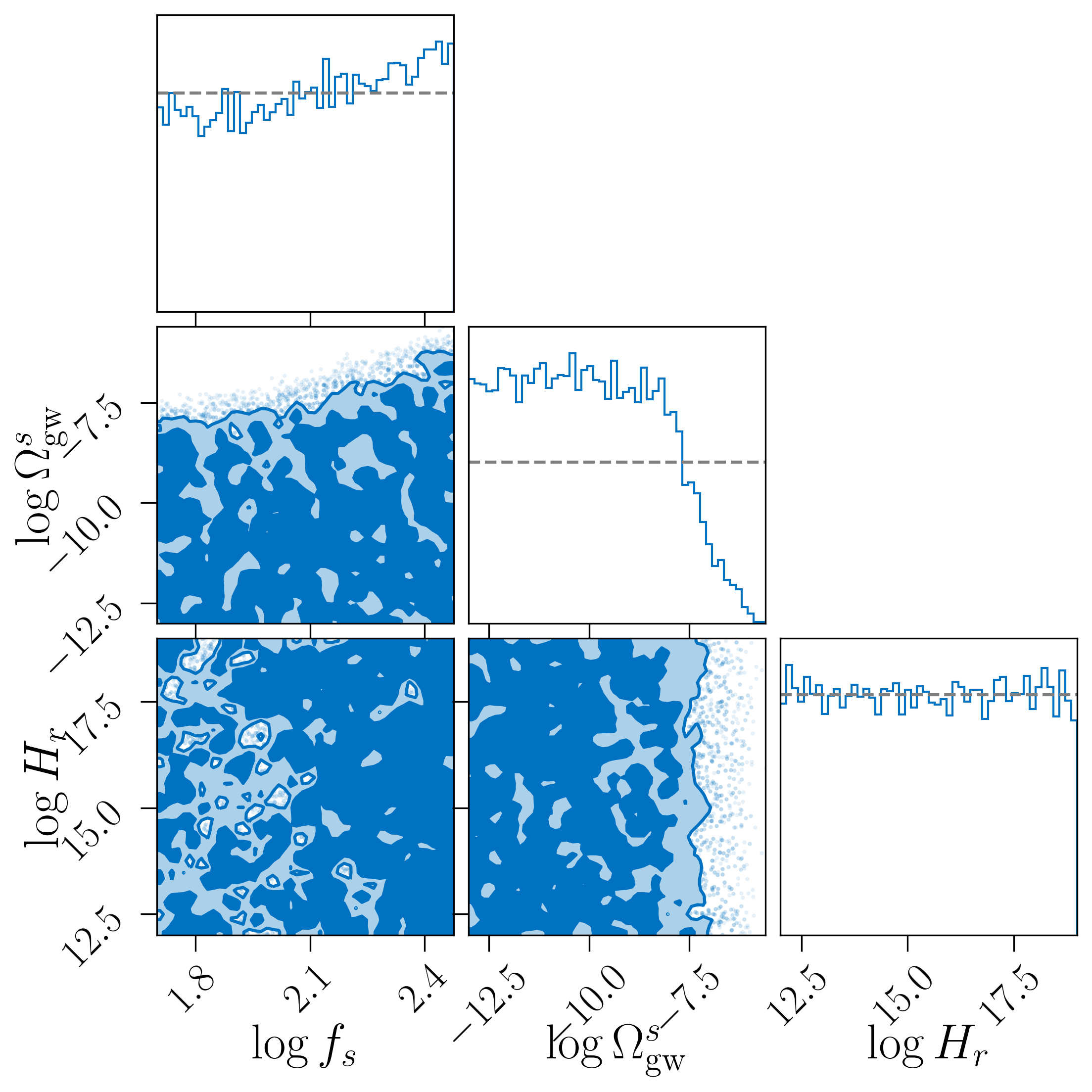}
    \caption{Posterior distributions of the parameters appears in the dilaton-string spectrum. The gray dashed lines denote the priors adopted. The exclusion contours at $68\%$ and $95\%$ CL are shown in blue shaded regions.} 
    \label{fig:posterior}
\end{figure}%
{\it Results.}
We present the posterior distributions of dilaton-string case in \Fig{fig:posterior}.
The Bayes factor $\log\mathcal{B}$ between model-noise and pure noise is $-0.16$, indicating that there is no such signal detected. At $95\%$ confidence level (CL), the upper limit of $\Omega_\text{gw}^s$ is between $10^{-8}\sim10^{-6.3}$, depending on the specific $f_s$. In addition, the result shows no preference for the Hubble parameter $H_r$ and its value do not obviously affect the distribution of remaining parameters.

\begin{figure}[ht]
    \centering
    \includegraphics[width=\columnwidth]{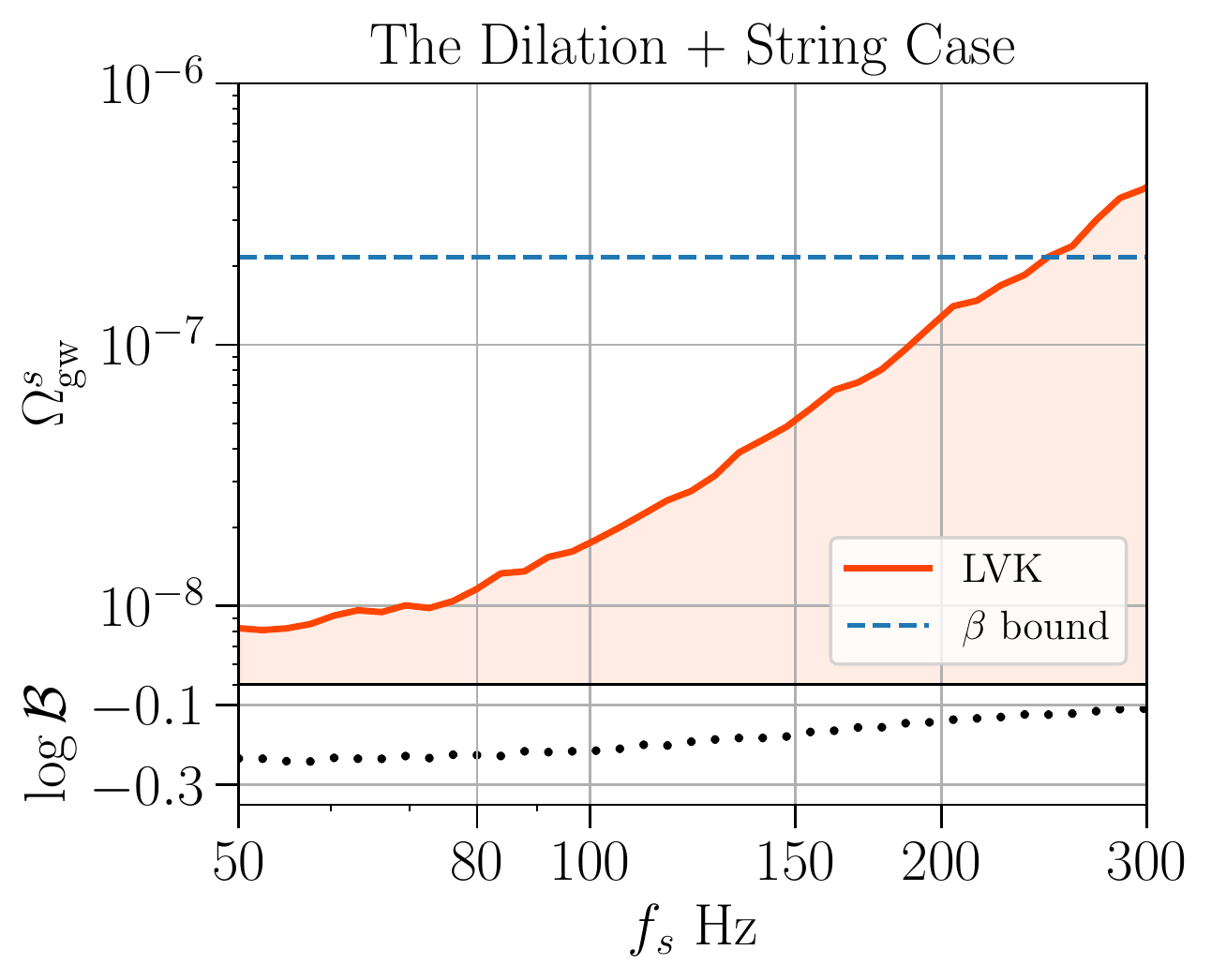}
    \caption{Upper limit of energy density $\Omega_\text{gw}^s$ produced in a dilaton-string case at $95\%$ CL from LVK. Besides, we denote the constraint from $\beta$ by blue dotted line. Bayes factors of the model with different $f_s$ are illustrated at the bottom.}
    \label{fig:dilstr}
\end{figure}
\begin{figure}[ht]
    \centering
    \includegraphics[width=\columnwidth]{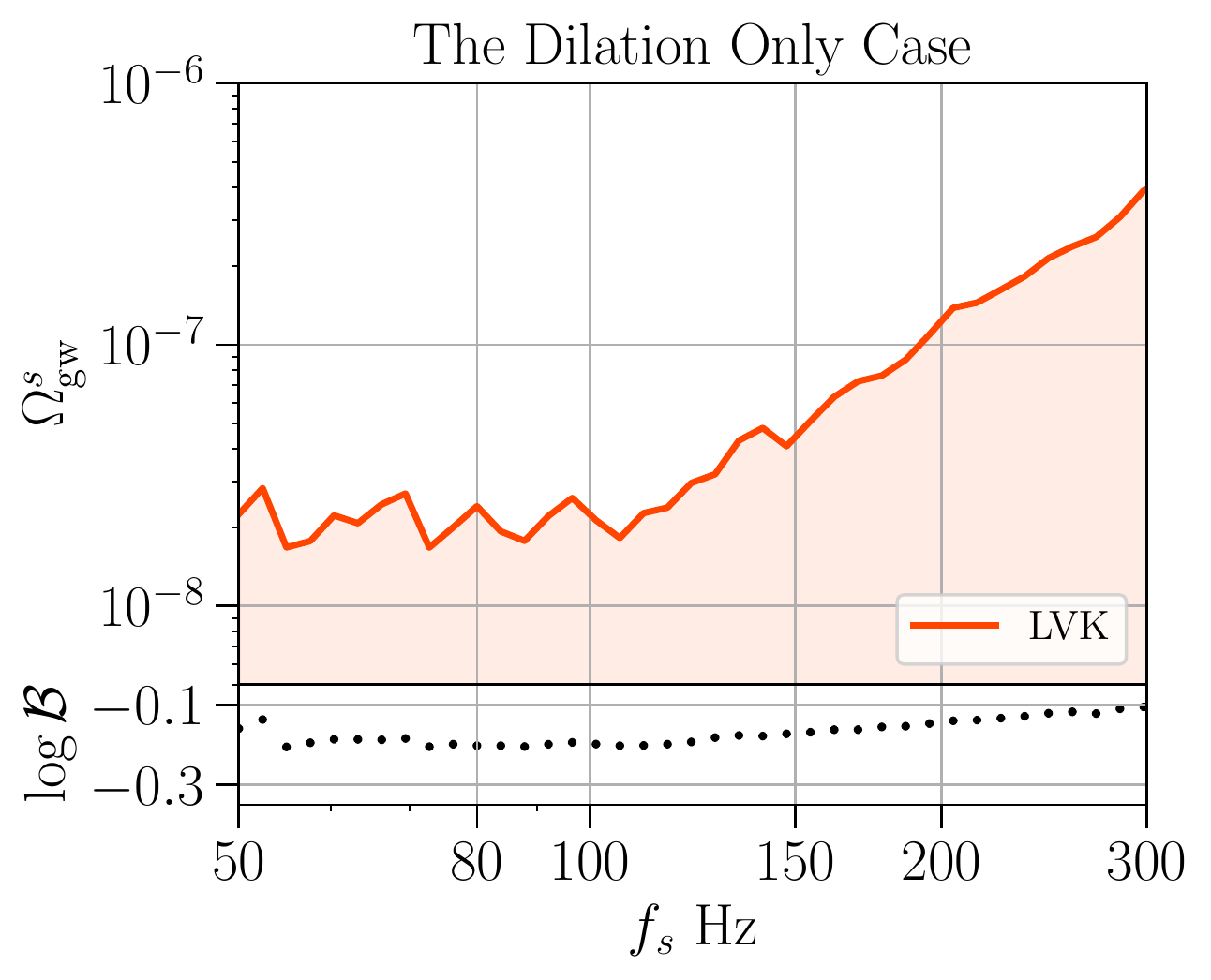}
    \caption{Upper limit of energy density $\Omega_\text{gw}^s$ produced in a dilaton only case at $95\%$ CL. Bayes factors of the model with different $f_s$ are illustrated at the bottom.}
    \label{fig:dilonly}
\end{figure}

%
Since there is no evidence of the existence of SGWB in the observing results, we choose to take $f_s$ as a series of values in the sensitive band of the ground detectors, and then attempt to obtain the upper limit of the fractional energy density $\Omega_\text{gw}^s$. This time we have assumed $H_r=5\times10^{17}\,\text{GeV}$. The dilation-string case and dilation only case are considered separately. In \Fig{fig:dilstr}, we present the upper limit of $\Omega_\text{gw}^s$ as a function of $f_s$ at $95\%$ CL for a dilaton-string spectrum. Parameter space above the curve is excluded. Notice that the power index $\beta$ should satisfy the constraint $\beta\geq0$ \cite{gasperini_2007}, this places an upper bound $\Omega_\text{gw}^s\leq2.2\times10^{-7}$ for the specific values of parameters we choose. The Bayes factor $\log\mathcal{B}$ varies from $-0.24$ to $-0.06$, showing that there is no such signal exists once again. Joint upper limits of $\Omega_\text{gw}^s$ from different $f_s$ lead to $\Omega_\text{gw}<1.7\times10^{-8}$ at $100\,\text{Hz}$. Similarly, we post the result of dilaton only case in \Fig{fig:dilonly}.
The Bayes factor is in the range of $-0.21\sim-0.06$ and $\Omega_\text{gw}(100\,\text{Hz})<2.1\times10^{-8}$.

\begin{figure}[ht]
    \centering
    \includegraphics[width=\columnwidth]{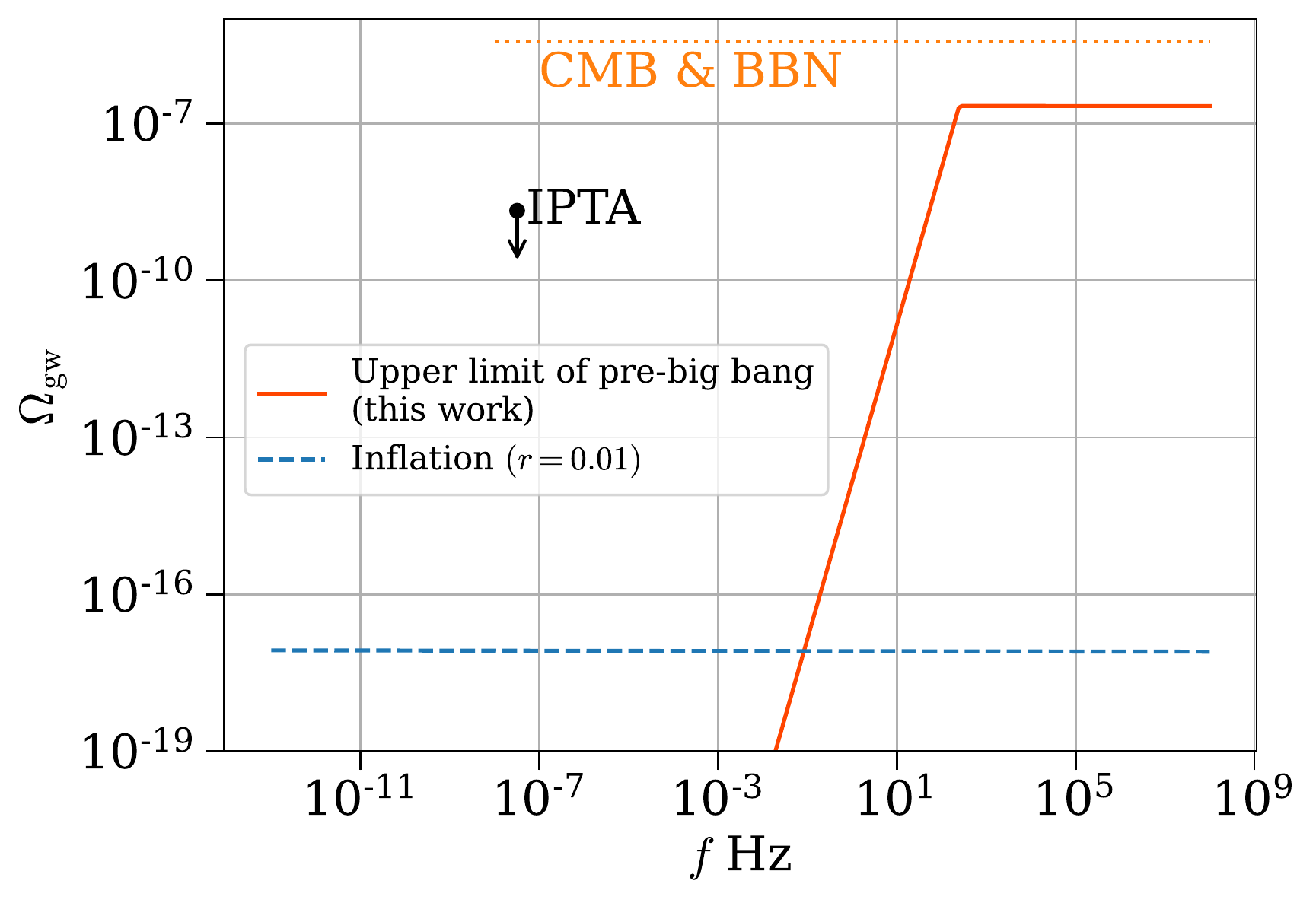}
    \caption{Upper limit on pre-big bang spectrum from LVK. The blue dashed line corresponds to the relic GW from inflation with tensor-to-scalar ratio $r=0.01$. The limits on SGWB from IPTA and indirect bound from CMB $\&$ BBN are also presented. Note that some  similar constraints on SGWB from other PTA observations are given in \cite{NANOGrav:2020bcs,Goncharov:2021oub,Chen:2021rqp,Antoniadis:2022pcn,Chen:2021wdo}. }
    \label{fig:spectrum}
\end{figure}

In order to compare our constraints with other observations, we illustrate the typical upper limit (e.g. $\beta=0$ and $f_s\simeq247$ Hz) of pre-big bang model in \Fig{fig:spectrum} where the prediction of relic GW from inflation with  tensor-to-scalar ratio $r=0.01$ corresponds to the blue dashed line \cite{Zhao:2013bba}. Compared to the CMB $\&$ BBN \cite{Pagano:2015hma} and international pulsar timing array (IPTA) \cite{Chen:2021ncc}, we find that our constraints on pre-big bang model are much more stringent. 


\section{Conclusion}
We have performed the search of SGWB generated by string cosmology models in LVK first three observing runs. By considering the spectra which satisfy $f_s\in [50, \,300]\,\text{Hz}$, we finally obtain $\Omega_\text{gw}<(1.7,\,2.1)\times10^{-8}$ at $100\,\text{Hz}$. This is the first time that we find the observing GW data can further rule out the parameter space restricted by the model itself due to the non-decreasing dilaton and stable cosmology background ($\beta$ bound).  Our research has shown the ability of ground detectors to explore the physics about string cosmology. As the detectors reach their final design sensitivity, a stochastic background with $\Omega_\text{gw}\geq10^{-9}$ at several hundred frequencies will expect to be detectable.

{\it Acknowledgments. } 
We acknowledge the use of HPC Cluster of ITP-CAS. This work is supported by the National Key Research and Development Program of China Grant No.2020YFC2201502, grants from NSFC (grant No. 11922303, 11975019, 11991052, 12047503), Key Research Program of Frontier Sciences, CAS, Grant NO. ZDBS-LY-7009, CAS Project for Young Scientists in Basic Research YSBR-006, the Key Research Program of the Chinese Academy of Sciences (Grant NO. XDPB15).

\bibliography{jcap_sub.bbl}
\end{document}